\documentclass{ws-ijmpa2}

\title{MEASURING THE EARTH-SUN DISTANCE DURING A LUNAR ECLIPSE}
\author{Costantino Sigismondi}
\address{Department of Physics, University of Rome "La Sapienza" and ICRA, 
International Center for Relativistic Astrophysics, P.le A.Moro 2 00185 
Rome Italy}
\address{In memory of my uncle Luigi Santi}
\date{\today}
\def\be{\begin{equation}}
\def\ee{\end{equation}}

\begin{document}
\maketitle
\begin{abstract}
The classical method for measure the Earth-Sun distance is due to Aristarchus and it is based upon the measure of the angle Moon-Earth-Sun when the Moon is exactly in quadrature. Such an angle is only 9 arcminutes smaller than 90 degrees, and it is very difficult to evaluate, being necessary to look directly towards the Sun.
The distance Earth-Moon and the Earth's diameter are necessary ingredients in order to derive the value of the astronomical unit. This method requires also the knowledge of the Moon's distance and the Earth's diameter, but it can permit a more precise measurement of the involved angles.
\end{abstract}

\section{The method}

\begin{figure}
\centerline{\psfig{file=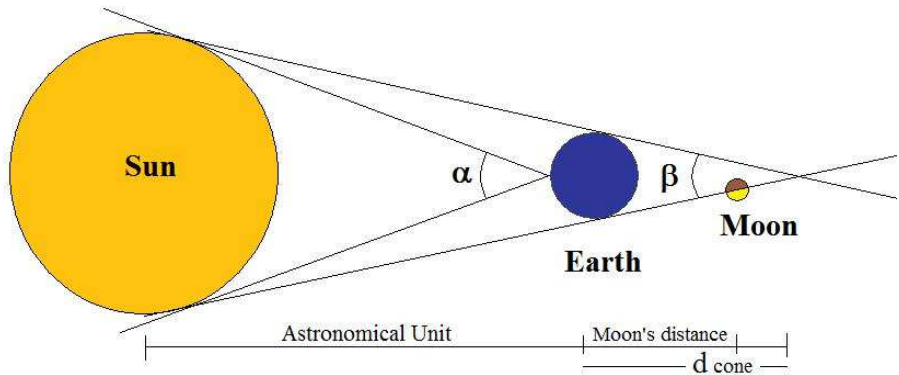,width=12cm}}
\vspace*{0pt}
\caption{The geometry of the Earth shadow during a lunar eclipse.}
\end{figure}

During a total eclipse of the moon, and even during a partial eclipse of magnitude something larger than 1/2, it is possible to observe the configuration shown in figure 2, where the shadow of the Earth crosses the full Moon along a diameter.
Taking a photo or projecting the image of the Moon with a pinhole\cite{sig} on a white paper it is possible to draw with a pencil the Earth's shadow's profile, in order to evaluate its diameter with respect to that of the Moon. 
The same geometry was utilized by Hipparcos for evaluating the position of the centre of the Earth shadow during the eclipse, studying the  precession of the equinoxes.\cite{bar}

From the distance DC expressed in term of the fraction $\frac{1}{a}$ of the Moon radius $r_{Moon}=EC$, we obtain that the Earth's shadow radius at the Moon's distance $R_{\oplus,M}$ is given by the equation:
\be
R_{\oplus,M} = r_{Moon} \cdot \frac{a^2+1}{2a}
\ee
at the Moon's distance.

Such value is to compare with the Earth's diameter, already known by Eratostenes,\cite{bar} in order to evaluate the angle with which the Sun is seen to appear as large as the Earth, as seen from the vertex of the shadow's cone. Also the position of that vertex is calculable.

\begin{figure}
\centerline{\psfig{file=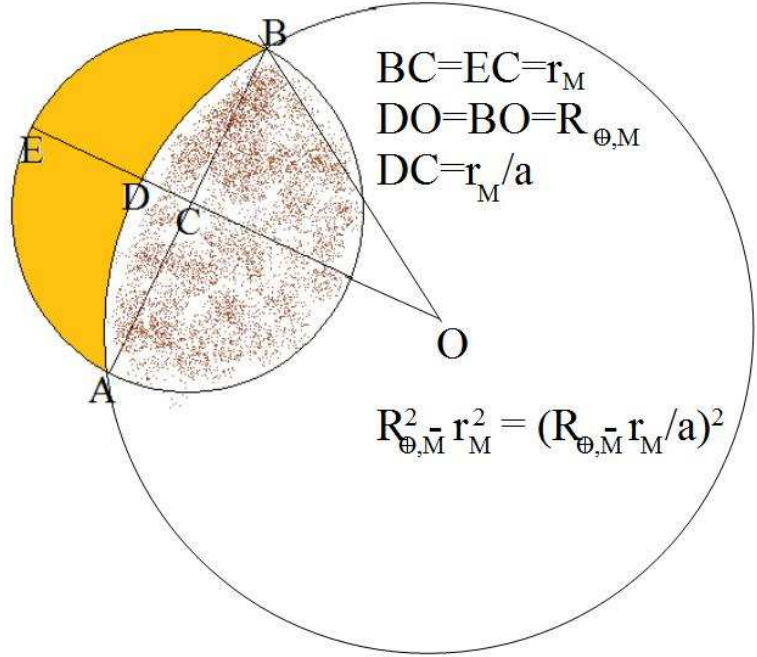,width=12cm}}
\vspace*{0pt}
\caption{The geometry of the Earth's shadow above the eclipsed Moon.}
\end{figure}

The position of the Sun is somewhere in the backwards extension of shadow's cone. It needs another constraint in order to fix its position. That is possible knowing its angular extension, a measure obtainable with a daily pinhole projection with a precision up to 30 arcseconds.\cite{sig}

\section{The errors of measure of the Astronomical Unit}
For the total eclipse of the Moon of January $21^{st}$, 2000, here it is calculated as exemplum the uncertainty of the AU evaluation. I observed the eclispe in Padua, Italy, with a 8x21 binoculars and I could roughly evaluate that $a \in [\frac{1}{5} \div \frac{1}{4}]$. By the formula (1) we obtain that $R_{\oplus, M} \in [2.80 \div 3.56]\times r_{Moon}$ at the Moon's distance. The Moon's distance at that moment (5 UT) was $d_{Moon}=360715~$Km from ephemerides.\cite{eff} Being the Earth's radius $6378$ Km, the Moon's radius $1737$ Km, 
the apparent diameter of the Sun $\alpha=1950$ arcsec, resolving the geometry of figure (1) with the equations
\be
d_{cone}=\frac{R_{\oplus}\cdot d_M}{R_{\oplus}-R_{\oplus,M}}=\frac{6378\cdot 360715}{6378-R_{\oplus,M}}\in [0.86 \div 1.24]\times 10^6  {\rm Km}
\ee
and 
\be
{\rm AU_{ev}}=\frac{d_{cone}\cdot \tan\beta}{\tan\alpha-\tan\beta}=\frac{R_{\oplus}}{\tan\alpha-(R_{\oplus}/d_{cone})}
\ee

with $d_{cone}$ the distance between the vertex of the cone of the Earth's shadow from the Earth's centre,
we obtain that our evaluation of the Astronomical Unit $AU_{ev}$ ranges as $AU_{ev} \in [1.5\div 3.2]\times 10^6$ Km.
The result of Aristarchus was $\frac{1}{20}$ of the true value,\cite{bar} and this method gives the same order of magnitude
of that one, with an estimation from $\frac{1}{50}$ to  $\frac{1}{100}$ of the true value.  
A concrete improving of such evaluation can be otained working with a pinhole apparatus, or, obviously, with the photography, being critical the evaluation of $d_{cone}$ in the denominator of the equation (3).
 
\section{Conclusions}

That method requires the same hypothesis of the Aristarchus' one,\cite{bar} and it requires instrumentations available even during classical epochs. So it can be considered equivalent to a classical method for measuring the Earth-Sun distance.
\section*{Acknowledgments}
Thanks to dr. Renato Klippert.

\end{document}